\documentclass[preprint,12pt]{elsarticle}

\usepackage{mathtools}
\usepackage{algorithm}
\usepackage[noend]{algpseudocode}
\usepackage{bm}
\usepackage{graphicx}
\usepackage[]{xcolor}

\makeatletter
\def\BState{\State\hskip-\ALG@thistlm}
\makeatother

\usepackage{amsmath}
\usepackage{amsfonts}
\usepackage{amssymb}
\usepackage{lineno}

\usepackage{upgreek}

\newcommand{\rmc}{{\mathrm c}}
\newcommand{\rmd}{{\mathrm d}}

\journal{Journal Name}

\begin{document}

\begin{frontmatter}

\title{Critical Exponent for the Lyapunov Exponent and Phase Transitions -- The Generalized Hamiltonian Mean-Field Model}

	\author{M.~F.~P.~Silva Jr.$^1$, T.~M.~Rocha Filho$^{1,2}$ and Y. Elskens$^3$}

	\address{${}^1$ Instituto de F{\'\i}sica, Universidade de Bras\'{\i}lia, Bras\'{\i}lia - Brazil.\\
	$^2$ International Center for Condensed Matter Physics, Universidade de Bras\'{\i}lia,\\ 
	         Bras\'{\i}lia - Brazil\\
	$^3$ Aix-Marseille Universit\'{e}, CNRS, UMR 7345 PIIM, \\
	        case 322 campus Saint-J\'er\^ome, F-13397 Marseille cedex 13 - France}

\begin{abstract}
We compute semi-analytic and numerical estimates for the largest Lyapunov exponent 
in a many-particle system with long-range interactions,
extending previous results for the Hamiltonian Mean Field model with a cosine potential.
Our results evidence a critical exponent associated to a power law decay of the largest Lyapunov exponent
close to second-order phase-transitions, close to the same value
as for the cosine Hamiltonian Mean Field model, suggesting the possible universality of this exponent.
We also show that the exponent for first-order phase transitions has a different value from both theoretical
and numerical estimates.
\end{abstract}


\end{frontmatter}

\section{Introduction}
\label{S:1}

The dynamics of classical many-body systems with long range interactions in a $D$-dimensional space, with potential decaying
at large distances $r$ as $r^{-D}$~\cite{campa2014physics},
are described exactly by a Vlasov equation, where a Kac prescription is used in order
to have a properly defined continuum limit~\cite{kac}.
In this limit, particles interact only through their mean field~\cite{braun,nosval,spohn,jabin14,kiessling14},
and the system never reaches thermodynamic equilibrium, and usually settles into a non-Gaussian stationary state~\cite{nosval}.
For a stationary state they effectively become pairwise uncoupled, like particles evolving in a static potential.
This implies that one-dimensional models with long-range interactions are integrable,
and therefore non-chaotic, in this limit. 
On the other hand, a more complex situation emerges for a finite number of particles, 
where collisional contributions~\cite{rouet1991relaxation,chaffi2017convergent,esceldo15coulomb,ebezd18revmodplaphy} 
become relevant to the dynamics, and correct the simple mean-field picture, usually implying chaos.
These collisional corrections are also responsible for driving the system towards thermodynamic equilibrium,
although with very long relaxation times~\cite{scaling,chris}.

To show that the system dynamics is chaotic amounts to show that its largest Lyapunov exponent (LLE) is positive~\cite{ott}, 
which has been used successfully
for long-range interacting systems~\cite{vallejos1,firpo1,anteneodo1,miranda2018lyapunov,luciano2}.
A geometrical approach based on statistical averages of microscopic dynamics 
was developed by Casetti and
collaborators~\cite{casetti1995gaussian,casetti2000geometric,casetti1996riemannian,caiani1997geometry}.
Firpo~\cite{firpo1998analytic} used this approach to show that, for the cosine Hamiltonian Mean-Field
model~\cite{antoni1995clustering}, the LLE $\lambda$ scales as $\lambda\propto | e-e_\rmc |^{1/6}$
at the second-order phase transition, 
with $e$ the system energy per particle and $e_\rmc$ its critical value.
This result was corroborated in Ref.~\cite{miranda2018lyapunov} from  molecular dynamics simulations,
although the values of the LLE obtained numerically deviate from the theoretical predictions in~\cite{firpo1998analytic}.

In the present work, we extend this analysis to the Generalized Hamiltonian Mean Field (GHMF)~\cite{teles2012nonequilibrium},
which has a richer phase diagram than the cosHMF model, with different second-order and also a first-order transitions.
This enables us to verify whether the scaling exponent for the LLE depends on the nature of the phase transition 
and whether its value is model dependent.

This paper is structured as follows: In Section~\ref{S:2} we present the GHMF model and its main properties.
In Section~\ref{S:3} we review the analytical and numerical approaches for the determination of the LLE.
Our main results are presented in Section~\ref{S:4} and we close the paper with some concluding remarks in Section~\ref{S:5}.

\section{Generalized Hamiltonian mean field model}
\label{S:2}

The model was introduced in~\cite{teles2012nonequilibrium} and consists of $N$ particles with position $\theta_i$
on a circle and conjugate momentum $p_i$, with the Hamiltonian
\begin{equation}
  H   = \sum_{i=1}^{N} \dfrac{p_i^2}{2} + \frac{1}{2N} \sum_{i,j = 1}^{N} v(\theta_i - \theta_j),
  \label{hamil-V}
\end{equation}
with the potential 
\begin{equation}
  v(\theta)
  = 1 - \Delta \cos\theta - (1 - \Delta) \cos (q\theta),
  \label{pot-V}
\end{equation}
where $q$ is a positive integer and $\Delta \in [0, 1]$.
The familiar cosine mean-field model is recovered with $\Delta = 1$.
The GHMF model is solvable at equilibrium and the numeric effort of Molecular Dynamics (MD) simulations
scale with $N$ instead of the usual $N^2$, which allows for large $N$ simulations~\cite{teles2012nonequilibrium,antoni1998sandoz}. 
By defining
\begin{equation}
  \mathbf{m}_1 
  = (m_{1x}, m_{1y}) = m_1 (\cos \varphi_1, \sin \varphi_1)
  = (\langle \cos \theta \rangle, \langle \sin \theta \rangle),
  \label{m1-def}
\end{equation}
and
\begin{equation}
  \mathbf{m}_q 
  = (m_{qx}, m_{qy}) = m_q (\cos \varphi_q, \sin \varphi_q)
   = (\langle \cos q\theta \rangle, \langle \sin q\theta \rangle),
  \label{mq-def}
\end{equation}
the Hamiltonian is rewritten as
\begin{equation}
  H 
  = \sum_{i=1}^{N} \frac{p_i^2}{2} + \frac{N}{2} \left[ 1 - \Delta \, m_1^2 - \left( 1 - \Delta \right) m_q^2 \right].
  \label{hamil-GHMF}
\end{equation}
For the present purposes, we restrict ourselves here to the case $q = 2$,
also considered in~\cite{teles2012nonequilibrium, levin2014nonequilibrium,pikovsky}, for which,
besides a paramagnetic ($m_1 = m_2 = 0$) and a ferromagnetic ($m_1 > 0,\ m_2 > 0$) phases, 
the model also presents a nematic ($m_2 > m_1 = 0$) phase. The transitions are second-order 
except for an interval of $\Delta$ values where the ferromagnetic-paramagnetic transition is first-order
(see Fig.~47 of Ref.~\cite{levin2014nonequilibrium}).


\section{Estimation of the largest Lyapunov exponent}
\label{S:3}

Let us consider the vector
\begin{equation}
  \bm{x}(t) 
  \equiv (x_1(t),x_2(t),\ldots,x_n(t)),
        \label{definicao-vetor}
\end{equation}
satisfying a set of $n$ first-order differential equations
\begin{equation}
  \dfrac{\rmd \bm{x} (t)}{\rmd t} 
  = \bm{F} (\bm{x} (t)).
    \label{sis-nlinear}
\end{equation}
The Lyapunov exponent is a measure of the growth rate of the difference vector $\bm{y}(t)$ between two neighbor
trajectories and given by
\begin{equation}
  \lambda 
  = \lim_{t \rightarrow \infty} \lim_{\vert \vert \bm{y}(0) \vert \vert \rightarrow 0}
	\dfrac{1}{t} \ln \dfrac{\vert \vert \bm{y} (t) \vert \vert}{\vert \vert \bm{y}(0) \vert \vert}.
	\label{definicao-lyapunov}
\end{equation}
The value of $\lambda$ usually depends on $\bm{y}(0)$, generating a Lyapunov spectrum:
\begin{equation}
  \lambda_1 \geq \lambda_2 \geq \dots \geq \lambda_n.
	\label{espectro de lyapunov}
\end{equation}
Since we are only interested here in $\lambda_1$, the largest of all such exponents, from now on
we drop the index in $\lambda_1$.

\subsection{Analytical estimation}
\label{S:31}

Casetti and collaborators~\cite{casetti1995gaussian,caiani1997geometry,pettinibook} 
develop an approach for the analytical estimation of the LLE from a geometrical approach for
the dynamics, based on Riemannian geometry, such that trajectories correspond to geodesics of an underlying metric.
Chaos then arises as instabilities in the flow of such geodesics, 
which depends on the properties of the curvature of the Riemannian manifold~\cite{casetti1996riemannian}.
Assuming that the effective fluctuations of the curvature along the trajectory are described by a Gaussian stochastic process, 
the LLE is given~\cite{caiani1997geometry} by
\begin{equation}
  \lambda 
  = \dfrac{\Lambda}{2} - \dfrac{2\kappa_0}{3\Lambda},
	\label{LLE-analitic}
\end{equation}
with
\begin{equation}
  \Lambda^3=2\sigma^2_{\kappa} \tau + \sqrt{\dfrac{64}{27}\kappa^3_0 + 4\sigma^4_{\kappa}\tau^2} ,
    \label{Lambda}
\end{equation}
and
\begin{equation}
  \tau= \dfrac{\pi \sqrt{\kappa_0}}{2\sqrt{\kappa_0}\sqrt{\kappa_0 + \sigma_{\kappa}} + \pi \sigma_{\kappa}},
	\label{tau}
\end{equation}
where $\kappa_0 \equiv \langle \kappa_R \rangle_{\mu}$ and $\sigma^2_{\kappa} \equiv \langle \delta^2 K_R \rangle_{\mu}$
(the $\langle\cdots\rangle_\mu$ stands for the microcanonical ensemble average)~\cite{casetti1996riemannian}.
The curvature $\kappa_R$ is given by the Laplacian of the total potential energy $V$ of the system,
\begin{equation}
  \kappa_R = \dfrac{K_R}{N-1} = \dfrac{\nabla^2 V}{N-1}.
  \label{Ricci-curvature}
\end{equation}

For the cosHMF model ($\Delta = 1$), Eq.~\eqref{Ricci-curvature} yields the estimate
\begin{equation}
  \kappa_0 
  = \langle \kappa_R \rangle_{\mu} = 1 - \dfrac{2}{N-1} \langle V \rangle_{\mu} .
  \label{kR}
\end{equation}
Using the results in Ref.~\cite{pearson1985laplace}
for the mean and variance of the potential energy in the microcanonical ensemble, 
the following expressions for $\kappa_0$ and $\sigma_{\kappa}$ are obtained:
\begin{eqnarray}
  \kappa_0 & = & \bar{m}^2,
  \label{kappa-HMF}
  \\
  \sigma_{\kappa} &=& T \sqrt{2 - \dfrac{1}{c}}, 
  \label{sigma-HMF}
\end{eqnarray}
where $\bar{m}$, $T$ and $c$ are the magnetization, temperature and specific heat in the microcanonical ensemble, respectively.
We discuss below how these same quantities can be determined for other values of $\Delta$.

\subsection{Numerical estimation}
\label{ss:TM}

We estimate the LLE from the Tangent Map (TM) method~\cite{benettin1976kolmogorov, parker2012practical},
which consists to simultaneously evolve the original non-linear equations in Eq.~(\ref{sis-nlinear}) 
and the linearized equations for the difference vector corresponding to a neighbor solution.
For that purpose, one considers two nearby solutions $\bm{x}_A(t)$ and $\bm{x}_B(t)$ and their difference vector
\begin{equation}
  \bm{w}(t) \equiv \bm{x}_A(t) - \bm{x}_B(t).
  \label{vetor-diferenca}
\end{equation}
The evolution of $\bm{w}(t)$ is then given at first order as
\begin{eqnarray}
	\frac{\rmd \bm{w}(t)}{\rmd t} & = & \bm{F}(\bm{x}_A(t)) - \bm{F}(\bm{x}_B(t))=
        \dfrac{\partial \bm{F}}{\partial \bm{x}} \Bigr|_{\bm{x} = \bm{x}_A}\cdot\bm{w}(t)
  \nonumber\\
	& \equiv & \bm{J}\bm{w}(t),
        \label{evolucao-w}
\end{eqnarray}
with $\bm{J}$ the Jacobian matrix of the vector field $\bm{F}$ along trajectory ${\bm{x}_A}(t)$.
For a Hamiltonian system with $N$ degrees of freedom, 
the Jacobian matrix has dimension $2N \times 2N$, and for a Hamiltonian of the form in Eq.~(\ref{hamil-V}) it
is given by
\begin{equation}
  J =
  \begin{pmatrix}
    0 & I\\
    \tilde{J} & 0
  \end{pmatrix},
        \label{jacobiana-hamiltoniana}
\end{equation}
where $I$ is the $N \times N$ unit matrix and $\tilde{J}$ is the Hessian matrix of the potential 
\begin{equation}
  \tilde{J}_{ij} 
  = - \dfrac{\partial^2 V}{\partial \theta_i \partial \theta_j}.
        \label{Jtil}
\end{equation}
For an initial difference vector ${\bf w}_0 \equiv {\bf w}(0)$ with $||{\bf w}(0)||=\epsilon\ll1$,

After a fixed integration time $T_{\mathrm{norm}}$, it evolves to ${\bf w}_1\equiv {\bf w}(T_{\mathrm{norm}})$. 
Then ${\bf w}_1$ is normalized to $\epsilon$, and the procedure is iterated, 
generating a sequence of difference vectors ${\bf w}_i$, $i=1,2,\ldots$ 
The LLE is then given by
\begin{equation}
	\lambda
	=
	\lim_{k\rightarrow\infty}\frac{1}{k T_{\mathrm{norm}}}\sum_{i=1}^k\ln\frac{||{\bf w}_i||}{\epsilon}.
	\label{numlle}
\end{equation}

For the GHMF model, the Hamilton equations are
\begin{eqnarray}
  \notag
  \dot{\theta_i} &=& p_i,
	\nonumber\\
  \notag
  \dot{p_i} &=& \Delta \left( m_{1y} \cos \theta_i - m_{1x} \sin \theta_i \right)
                \\
                & & +\ (1 - \Delta) q \left[ m_{qy} \cos (q \theta_i) - m_{qx} \sin (q \theta_i) \right].
   \label{dinamica-ghmf}
\end{eqnarray}
The linearized equations for $\theta_i^\prime(t)=\theta_i(t)+\delta\theta_i(t)$ and $p_i^\prime(t)=p_i(t)+\delta p_i(t)$
around a solution of Eq.~(\ref{dinamica-ghmf}) are then
\begin{eqnarray}
  \notag
  \dot{\delta \theta_i} &=& \delta p_i,
	\nonumber\\
  \dot{\delta p_i} &=& \Delta \left( \delta m_{1y} \cos \theta_i^* - \delta m_{1x} \sin \theta_i^* \right)
  \nonumber\\
                           & & -\ \Delta \left( m_{1y}^* \sin \theta_i^* + m_{1x}^* \cos \theta_i^* \right) \delta \theta_i
			   \nonumber\\
                           & & +\ (1 - \Delta) q \left[ \delta m_{qy} \cos (q \theta_i^*) - \delta m_{qx} \sin (q \theta_i^*) \right]
			   \nonumber\\
               & & -\ (1 - \Delta) q^2 \left[ m_{qy}^* \sin (q \theta_i^*) + m_{qx}^* \cos (q \theta_i^*) \right] \delta \theta_i,
   \label{dinamicalinear-ghmf}
\end{eqnarray}
where
\begin{eqnarray}
\notag
\delta m_{1x} &\equiv& - \dfrac{1}{N} \sum_{j=1}^N \delta \theta_j \sin \theta_j^*,\\
\notag
\delta m_{1y} &\equiv& \dfrac{1}{N} \sum_{j=1}^N \delta \theta_j \cos \theta_j^*,\\
\notag
\delta m_{qx} &\equiv& - \dfrac{q}{N} \sum_{j=1}^N \delta \theta_j \sin (q \theta_j^*),\\
\delta m_{qy} &\equiv& \dfrac{q}{N} \sum_{j=1}^N \delta \theta_j \cos (q \theta_j^*).
  \label{delta-ms}
\end{eqnarray}

\begin{figure}[ht]
  \centering\includegraphics[width=0.9\linewidth]{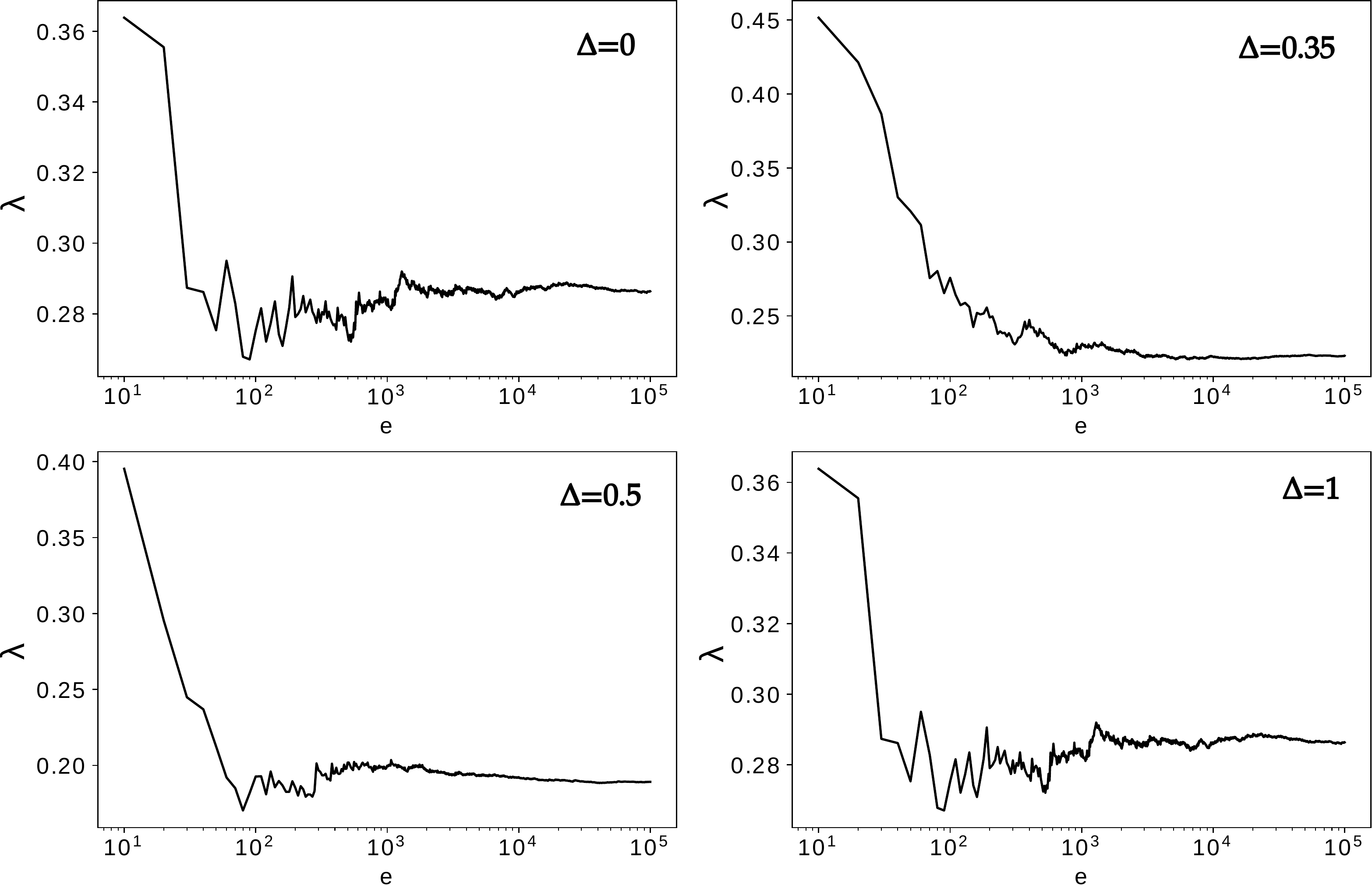}
  \caption{LLE for $N = 10^5$, $e = 0.5$ and $\Delta = 0$ (upper left), $\Delta = 0.35$ (upper right), 
                 $\Delta = 0.5$ (bottom left) and $\Delta = 1$ (bottom right) as a function of the integration time. 
	Integration time step is $\Delta t = 0.05$.}
\label{fig:convergencia}
\end{figure}

This approach was implemented in a parallel code on GPU, to compute the LLE for large values of $N$~\citep{miranda2018lyapunov, rocha2014molecular}.
Figure~\ref{fig:convergencia} shows the results for $N = 10^5$, $e = 0.5$ and $\Delta = 0, 0.35, 0.5, 1$, with a good convergence obtained for
total integration time $t_{\rm f} = 10^5$.

\section{Results}
\label{S:4}

\subsection{Theoretical predictions}

Following the prescription in Sec.~\ref{S:31}, we obtain the curvature $\kappa_R$ from
Eqs.~(\ref{Ricci-curvature}) and~(\ref{dinamica-ghmf}) as
\begin{eqnarray}
\notag
	\kappa_R & = &\frac{1}{N-1}\nabla^2 V=
	\frac{1}{N-1}\sum_i \dfrac{\partial}{\partial \theta_i} \dfrac{\partial V}{\partial \theta_i}=
	- \frac{1}{N-1}\sum_i \dfrac{\partial \dot{p_i}}{\partial \theta_i}
	\nonumber\\
	&=& \dfrac{1}{N(N-1)} \sum_{\substack{i,j=1\\(i\neq j)}}^N
	\left[ \Delta \cos(\theta_i - \theta_j) + (1-\Delta) q^2 \cos(q\theta_i - q\theta_j) \right]
	\nonumber\\
	&=& \frac{N}{N-1}\left[ \Delta m_1^2+(1-\Delta) q^2 m_q^2\right],
	\label{nablaV}
\end{eqnarray}
where in the last line we discarded a small term of order $1/N$. Thence we obtain for large $N$
\begin{equation}
\kappa_0 = \Delta m_1^2 + \left( 1 - \Delta \right) q^2 m_q^2.
	\label{kappa-GHMF}
\end{equation}
The results in Ref.~\cite{firpo1998analytic} for the cosHMF model are fully recovered by plugging $\Delta=1$ in Eq.~(\ref{kappa-GHMF})
and in the results below.
At variance with the results for the single cosine HMF, 
the right-hand side of Eq.~(\ref{kappa-GHMF}) is not a function of the potential energy of the system
$V=(N/2) \left[ 1 - \Delta \, m_1^2 - \left( 1 - \Delta \right) m_q^2 \right]$
 (due the $q^2$ multiplying the second term),
except for the cases $\Delta = 0$ and $\Delta = 1$, for which  
$m_q^{(\Delta = 0)} = m_1^{(\Delta = 1)}$ and $T^{(\Delta = 0)} = T^{(\Delta = 1)}$. 
Let us first consider these cases. We have that
\begin{eqnarray}
\notag
\kappa_0^{(\Delta = 0)} &=& q^2 \kappa_0^{(\Delta = 1)},\\
\sigma_{\kappa}^{(\Delta = 0)} &=& q^2 \sigma_{\kappa}^{(\Delta = 1)},
	\label{k0=k1}
\end{eqnarray}
that imply
\begin{equation}
  \lambda^{(\Delta = 0)} 
  = q \lambda^{(\Delta = 1)}.
  \label{l0=ql1}  
\end{equation}
Assuming now that, near the phase transition, 
$\lambda^{(\Delta = 0)}$ and $\lambda^{(\Delta = 1)}$ obey a scaling law of the form
\begin{eqnarray}
\notag
  \lambda^{(\Delta = 0)} &\propto& | e - e_\rmc |^{\xi_0},
  \\
  \lambda^{(\Delta = 1)} &\propto& | e - e_\rmc |^{\xi_1},
  \label{lambdas-ec}
\end{eqnarray}
with $\xi_0$ and $\xi_1$ the critical exponents for $\Delta=0$ and $\Delta=1$, 
respectively, we obtain from Eq.~\eqref{l0=ql1} that
\begin{equation}
  \xi_0 = \xi_1  .
  \label{xi0=xi1}
\end{equation}

For $0 < \Delta < 1$, the average curvature $\kappa_0$ is not a function of the potential energy alone, and
consequently the approach in~\cite{pearson1985laplace} cannot be extended directly to the present case.
In order to overcome this difficulty, we determine the LLE from Eq.~(\ref{LLE-analitic}),
with $\kappa_0$ given by Eq.~(\ref{kappa-GHMF}) 
and computing $\sigma_{\kappa}$ from a Microcanonical Monte Carlo (MMC) simulation.
Averages are computed by sampling equilibrium configurations ${\bm{\theta}} = ( \theta_1,\ldots,\theta_N )$ of the system, 
with acceptance probability~\cite{ray1991microcanonical}:
\begin{equation}
	P({\bm{\theta}} \rightarrow {\bm{\theta}}^\prime) 
	= {\rm min} \left( 1,\dfrac{W_E({\bm{\theta}})}{W_E({\bm{\theta}}^\prime)} \right),
	\label{prob-ac}
\end{equation}
where
\begin{equation}
	W_E({\bm{\theta}}) 
	\equiv (E-V({\bm{\theta}}))^{\frac{N}{2}-1} \, \Theta (E-V({\bm{\theta}})).
	\label{defW}
\end{equation}
with $E$ the energy of the system and $\Theta(\,\cdot\,)$ the Heaviside function.
Thus, with Eq.~(\ref{prob-ac}), samples of ${\bm{\theta}}$ are generated 
with distribution proportional to the microcanonical probability density.
After convergence is reached, the equilibrium state is sampled by the rule in Eq.~(\ref{defW})
and microcanonical averages can be computed. This enables us to obtain $\sigma_{\kappa}^2$ as
\begin{equation}
\sigma_{\kappa}^2 = N(\langle \kappa_R^2 \rangle_{\mu} - \langle \kappa_R \rangle_{\mu}^2).
	\label{sigma-mmc}
\end{equation}

\begin{figure}[ht]
\centering\includegraphics[width=0.9\linewidth]{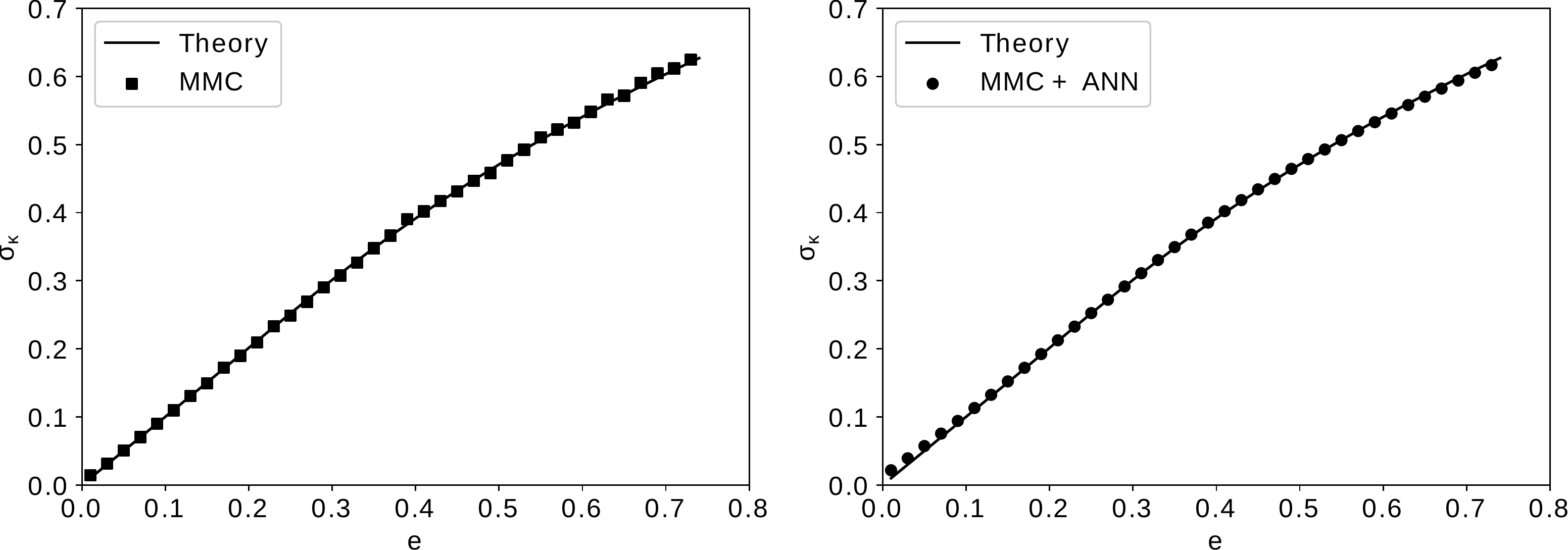}
\caption{Left panel: Values of $\sigma_{\kappa}$ from the MMC simulation.
	Right panel: The same results smoothed as described in the text.}
\label{fig:sigma-ray-ann}
\end{figure}

Numerical errors in this simulation, which propagate to the final value for the LLE,  
are particularly important near a phase transition.
To circumvent this limitation, we apply a nonlinear regression for the MMC results. 
For that purpose, we use a feedforward artificial neural network~\cite{mehta} 
(with 4 hidden layers of 32 neurons each and an exponential linear unit as activation function~\cite{clevert2015fast}).
The validity of this approach is evidenced in Fig.~\ref{fig:sigma-ray-ann} showing the results
from the smoothing procedure for $\Delta = 1$ compared to the corresponding theoretical prediction. 
The smoothing from the neural network correctly reproduces the values of $\kappa_0$ 
while reducing oscillations due to numerical errors.

\begin{figure}[ht]
  \centering\includegraphics[width=0.9\linewidth]{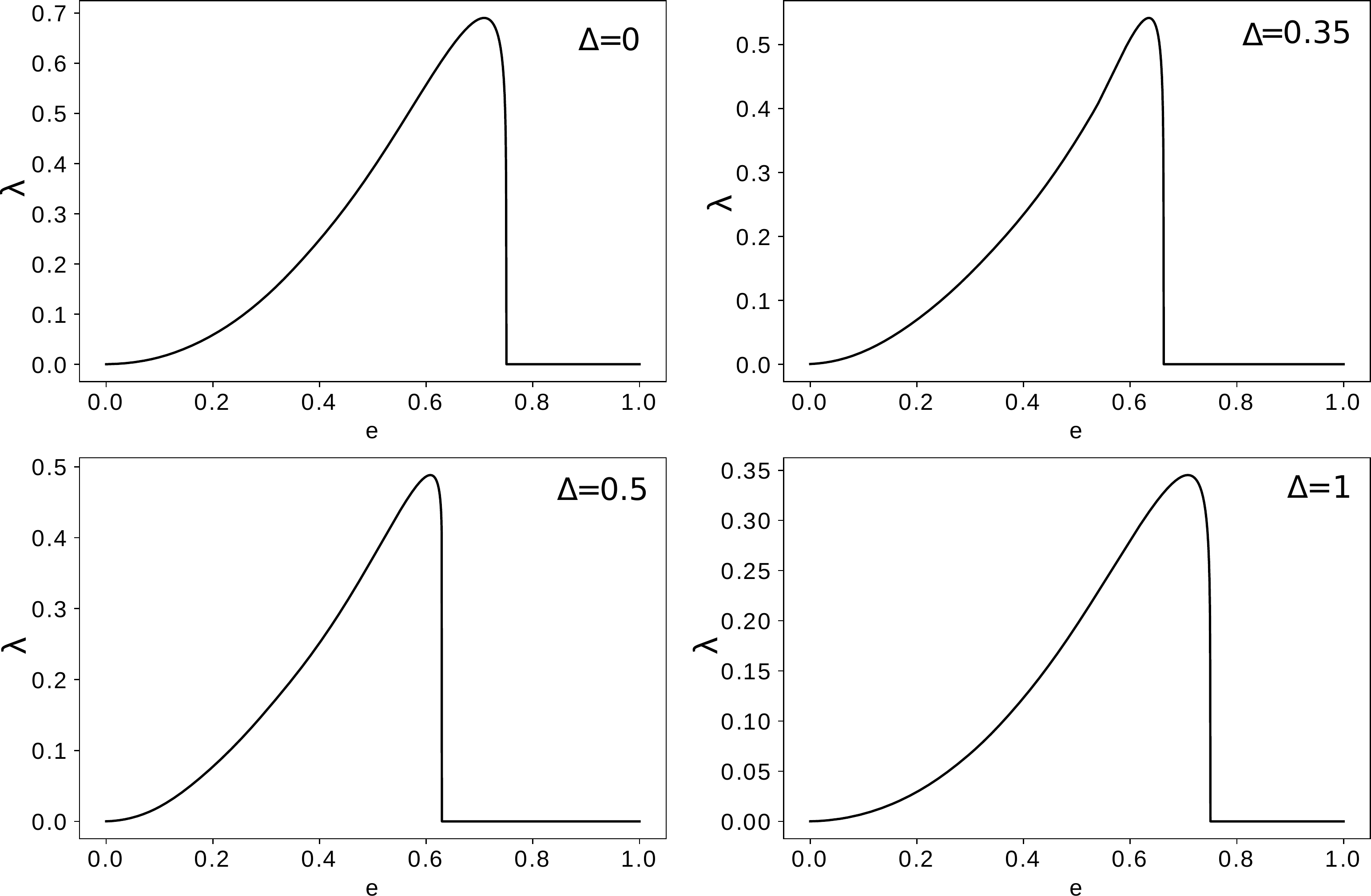}
  \caption{LLE for $\Delta = 0$ (upper left), $\Delta = 0.35$ (upper right),
                $\Delta = 0.5$ (bottom left) and $\Delta = 1$ (bottom right) as function of energy.}
  \label{fig:l-thermo}
\end{figure}

\begin{figure}[ht]
  \centering\includegraphics[width=0.9\linewidth]{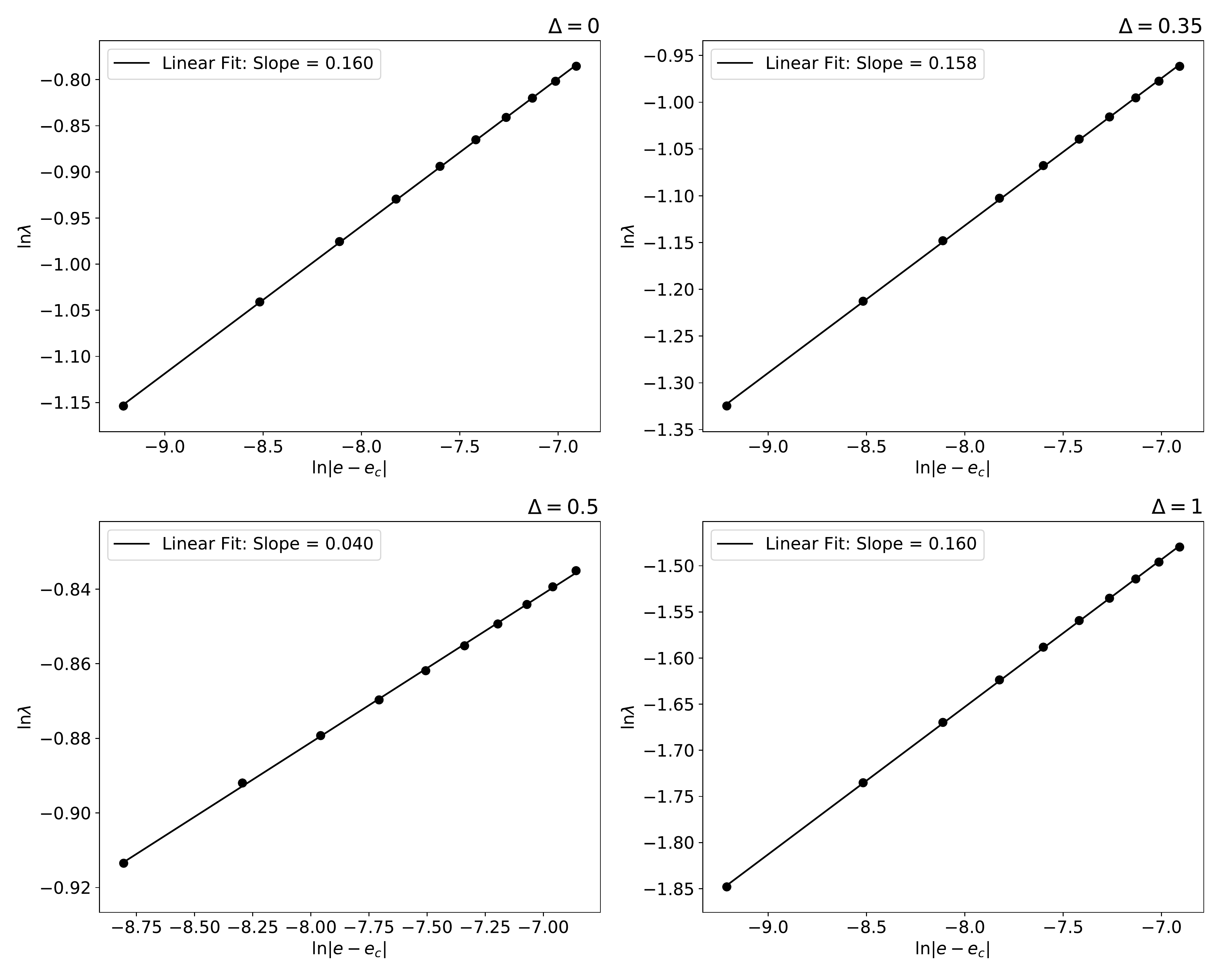}
  \caption{LLE close to the phase transition for $\Delta = 0$ (upper left), $\Delta = 0.35$ (upper right),
               $\Delta = 0.5$ (bottom left) and $\Delta = 1$ (bottom right).
               These results show that $\lambda$ has a critical exponent close to 1/6 when the transition is second order,
               but a different value when the transition is first order.}
  \label{fig:ec-thermo}
\end{figure}

\begin{figure}[ht]
\centering\includegraphics[width=0.9\linewidth]{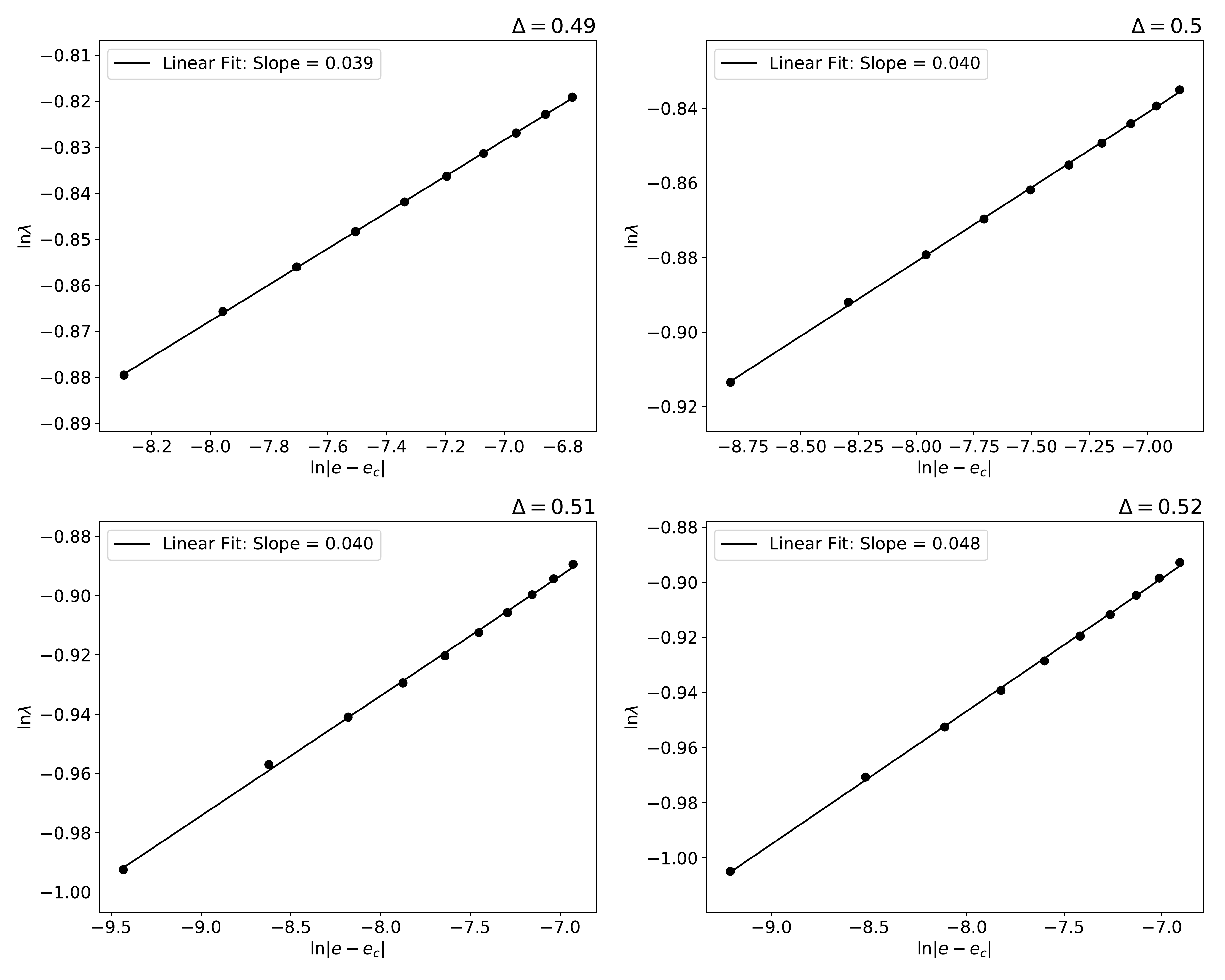}
\caption{Analytic estimates of $\lambda$ close to the phase transition for $\Delta = 0.49$ (upper left),
        $\Delta = 0.5$ (upper right), $\Delta = 0.51$ (bottom left) and $\Delta = 0.52$ (bottom right).}
\label{fig:ec-1order-thermo}
\end{figure}

Following the above prescriptions, the results obtained for the LLE as a function
of the energy per particle $e$, for $q = 2$ and a few values of $\Delta$,
are shown in Fig.~\ref{fig:l-thermo}.
We note that these results are in agreement with Eq.~\eqref{l0=ql1}. 
In order to investigate the scaling form $\lambda\propto|e-e_\rmc|^\xi$, we writea
\begin{equation}
  \ln \lambda = \xi \ln | e - e_\rmc | + C,
\end{equation}
with $C$ a constant. Figure~\ref{fig:ec-thermo} shows the log-log plots for the same cases as in Fig.~\ref{fig:l-thermo} near the phase transition,
and the corresponding values for the LLE. As a simple consistency test, we observe that Eq.~\eqref{xi0=xi1} is satisfied. 
More importantly, the critical exponents associated to second order transitions ($\Delta = 0,\ 0.35,\ 1$) 
are all very close to the value $1/6$ predicted from the analytical estimates. 
For the case $\Delta = 0.5$ with a first-order transition, the LLE also obeys a similar power law,
but with a different exponent $\xi=0.040$. Analytical estimates near the value $\Delta=0.5$ 
are shown in Fig.~\ref{fig:ec-1order-thermo}, with values for the power law exponent varying from $0.039$ to $0.048$,
far from the value for the continuous transition.

\begin{figure}[ht]
\centering\includegraphics[width=0.7\linewidth]{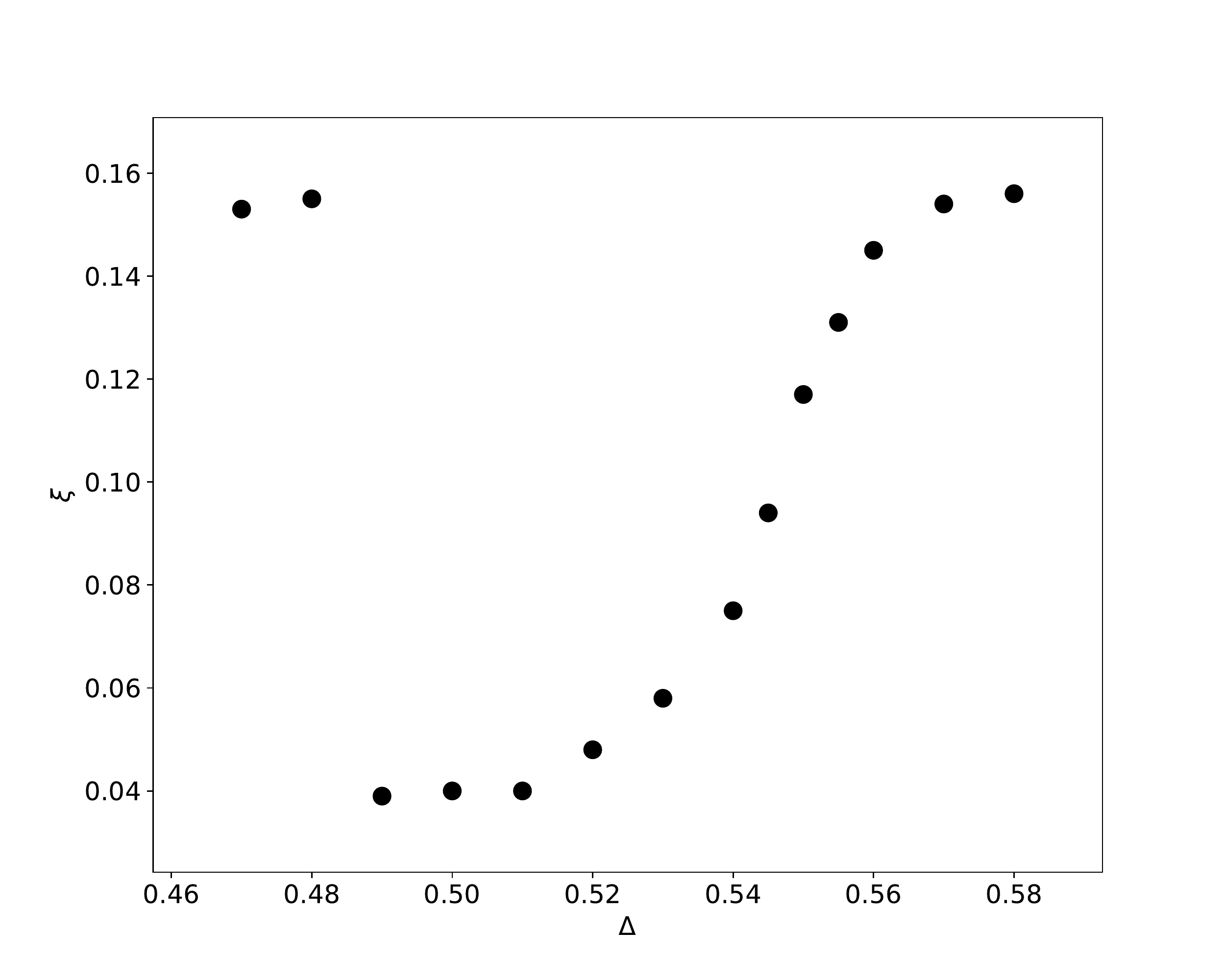}
\caption{Critical exponent of the LLE for values of $\Delta$ around the region of first-order transitions.}
\label{fig:beta-vs-delta}
\end{figure}

We also investigated the critical exponent $\xi$ by varying $\Delta$ in order to assess how
it changes when going from a second to a first order transition.  
The two tricritical points~\cite{pikovsky} occur at $\Delta \approx 0.545, e \approx 0.636$ and $\Delta \approx 0.477, e \approx 0.628$, 
and a critical end point occurs at $\Delta \approx 0.487, e \approx 0.628$.
Figure~\ref{fig:beta-vs-delta} shows that when increasing $\Delta$ starting from $\Delta=0.46$,
the value of the critical exponent drops abruptly from $\xi \approx 0.160$, related to second-order transitions, 
to $\xi \approx 0.040$, then it grows (smoothly but not linearly) to $0.160$ again,
but slightly departs from this value for the second-order transitions when $\Delta \gtrsim 0.545$.

\subsection{Molecular dynamics and the tangent map method}

\begin{figure}[ht]
  \centering\includegraphics[width=0.9\linewidth]{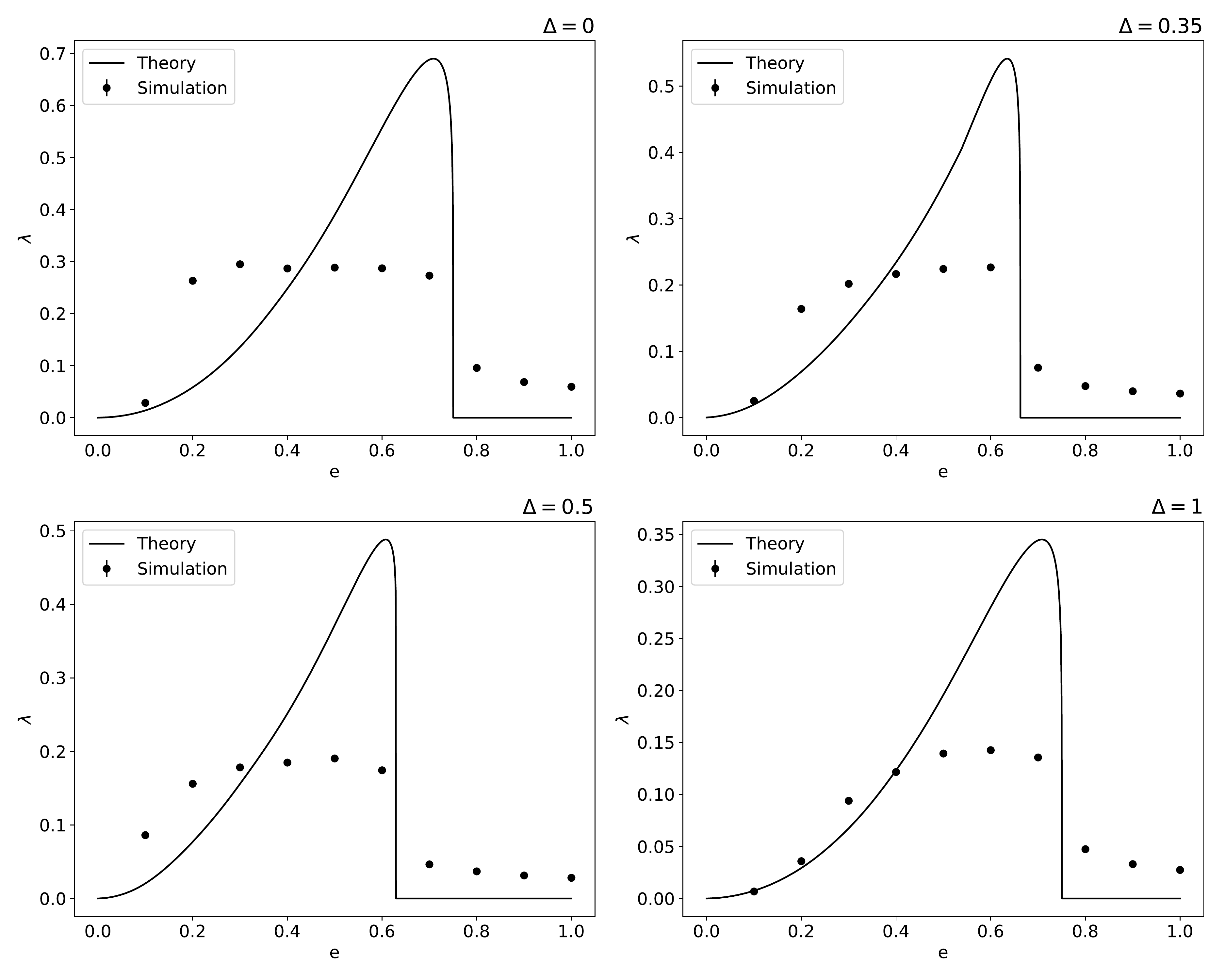}
  \caption{LLE from the tangent map method with $N = 10^4, 10^5, 10^6$ 
     and the corresponding theoretical predictions for $\Delta = 0$ (upper left),
	$\Delta = 0.35$ (upper right), $\Delta = 0.5$ (bottom left) and $\Delta = 1$ (bottom right) as a function of energy. The error bars are smaller than the symbol size.}
  \label{fig:l-thermo-dynamic}
\end{figure}

As systems with long-range interaction have a very long relaxation time to
equilibrium~\cite{campa2014physics,antoni1995clustering,teles2012nonequilibrium,levin2014nonequilibrium},
and since we know the analytic expression for the one-particle distribution at equilibrium,
we chose initial conditions for the MD simulation at the equilibrium state. 
The system is left to evolve for a time interval $t_0$ in order to thermalize before computing quantities of interest. 
To ensure that the system is indeed in the correct equilibrium state, we compute known thermodynamic properties
such as kinetic and potential energies and total magnetization, 
and check these values with respect to theoretical predictions.
The results for the TM method were obtained for $N = 10^4,\ 10^5$ and $10^6$, 
$\epsilon = 10^{-6}$ and total simulation time $T_{\mathrm f} = 10^5$,
with normalization of the difference vector $\bf{w}$ at time intervals of $T_{\mathrm{norm}} = 10$.
All results were checked for proper convergence.
The LLE as a function of energy per particle for the same parameter values considered in the previous section 
are shown in Fig.~\ref{fig:l-thermo-dynamic}, alongside the corresponding analytical estimates.

\begin{figure}[ht]
\centering\includegraphics[width=0.9\linewidth]{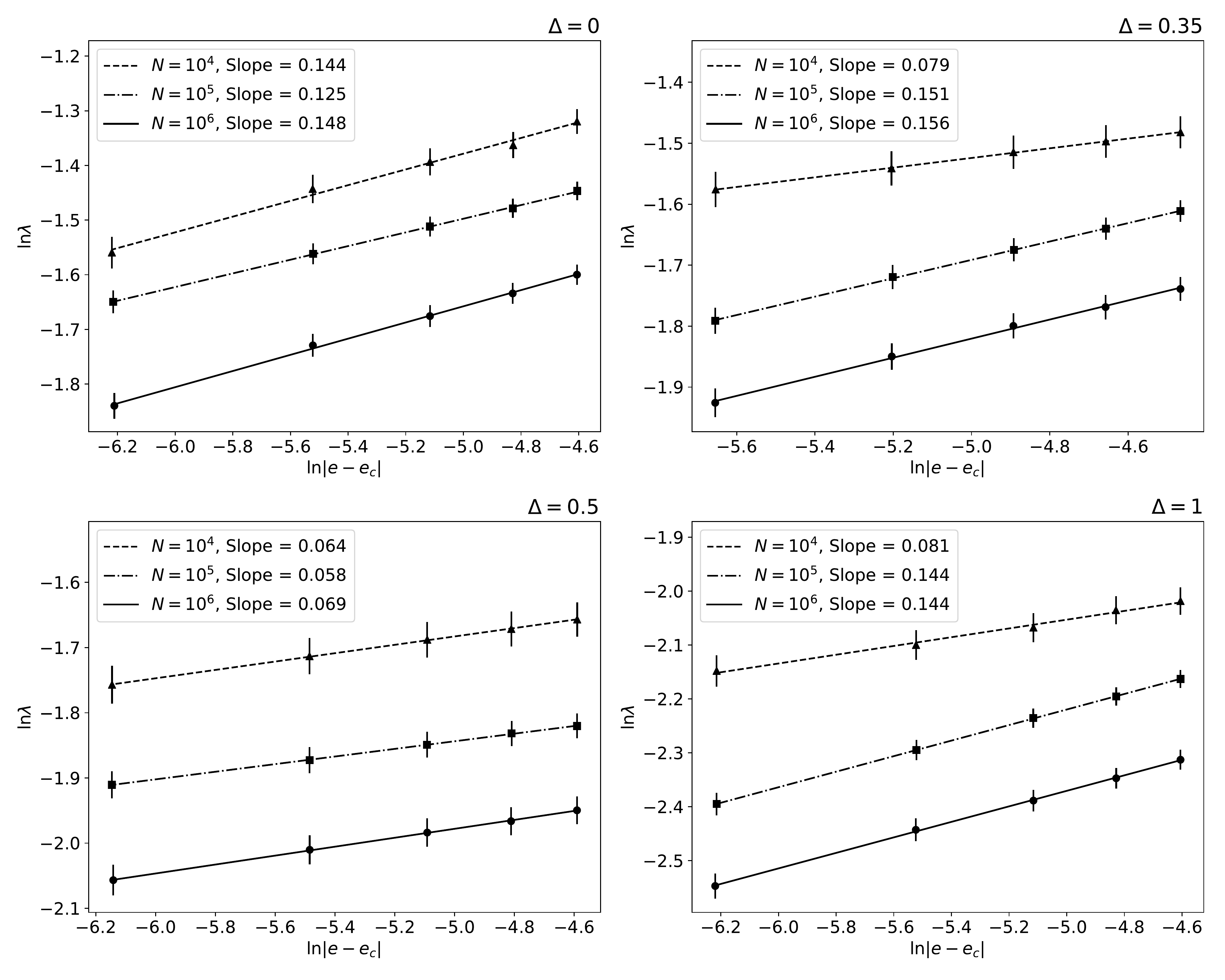}
\caption{LLE from the TM method for $N = 10^4, 10^5, 10^6$ close to the phase transitions for $\Delta = 0$ (upper left),
	$\Delta = 0.35$ (upper right), $\Delta = 0.5$ (bottom left) and $\Delta = 1$ (bottom right).}
\label{fig:ec-dynamic}
\end{figure}

Similarly to what was shown for the HMF model in Ref.~\cite{miranda2018lyapunov}, 
the LLE dependency on the energy differs significantly from its (semi-)analytic estimate. 
A possible explanation comes from the fact that one of the assumptions used in the analytical approach
is that fluctuations are $\delta$-correlated, which is expected to be valid only at higher
energies~\cite{miranda2018lyapunov,casetti1995gaussian,casetti2000geometric}. 
Figure~\ref{fig:ec-dynamic} shows the behavior of the LLE close to the critical energy for $N=10^4$, $N=10^5$
and $N=10^6$, and the value of the power law exponent $\xi$ from a least squares fit.
These results and the corresponding analytic predictions are summarized in Table~\ref{tab:ec}.
As the number of particles increases, the numeric estimates approach the analytical estimate,
except for the first-order phase transition.
For $\Delta = 0$ and $\Delta = 1$ (both for a second-order transition), the dynamical estimate is approximately
$10\%$ smaller than the analytical estimate, while it is only about $1\%$ smaller for $\Delta = 0.35$.

\begin{table}[H]
\centering
\begin{tabular}{ccccc}
\hline \hline
$\Delta$ & Theory & $N = 10^4$ & $N = 10^5$ & $N = 10^6$ \\ \hline
0        & 0.160  & 0.144  & 0.125  & 0.148 \\
0.35     & 0.158  & 0.079  & 0.151  & 0.156 \\
0.5      & 0.040  & 0.064  & 0.058  & 0.069 \\
1        & 0.160  & 0.081  & 0.144  & 0.144 \\ \hline \hline
\end{tabular}
\caption{Critical exponents obtained theoretically and dynamically.}
\label{tab:ec}
\end{table}

\begin{figure}[ht]
\centering\includegraphics[width=0.9\linewidth]{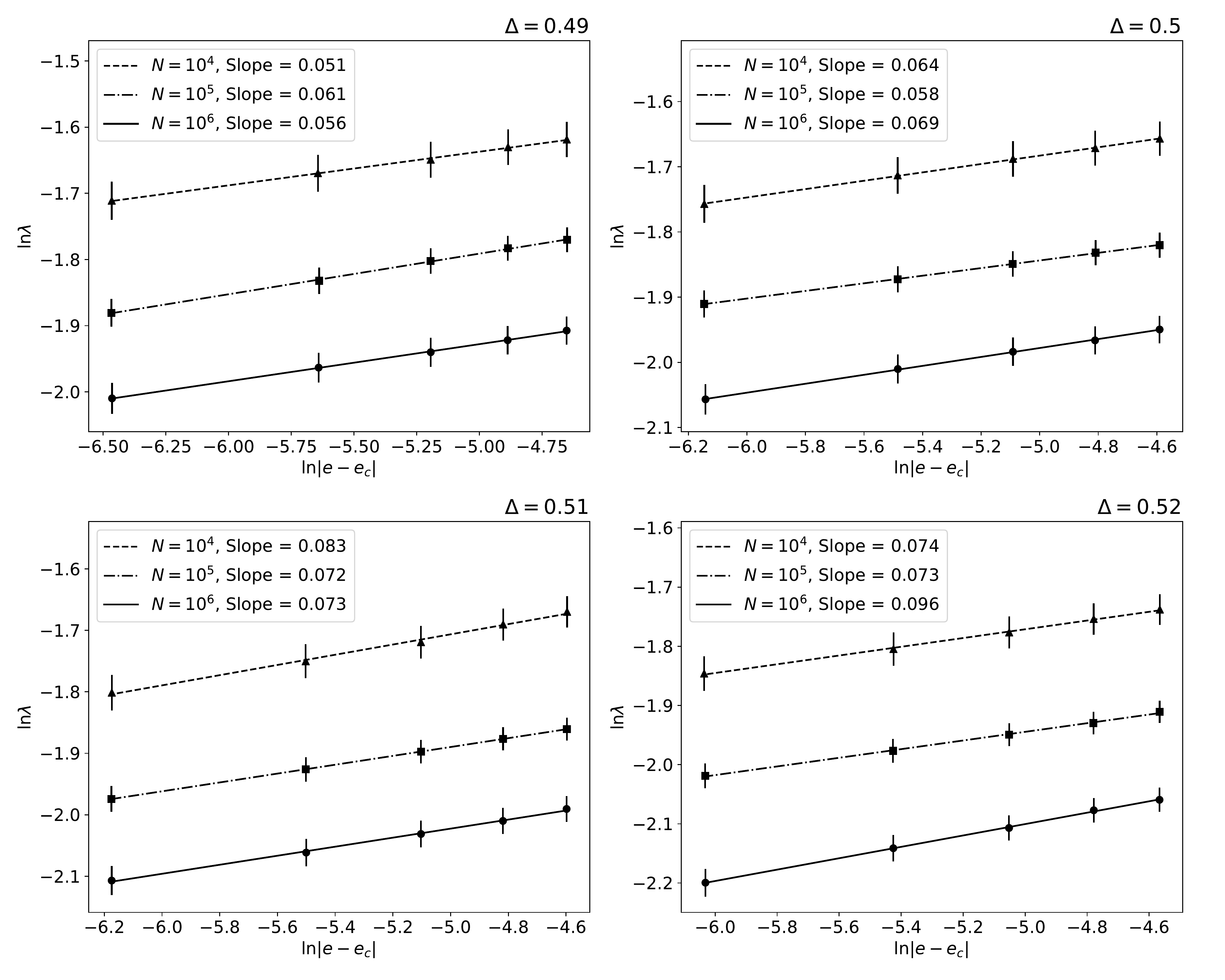}
\caption{LLE close to the phase transition from the TM method for $N = 10^4, 10^5, 10^6$ for $\Delta = 0.49$ (upper left),
	$\Delta = 0.5$ (upper right), $\Delta = 0.51$ (bottom left) and $\Delta = 0.52$ (bottom right).}
\label{fig:ec-1order-dynamic}
\end{figure}

We look now more carefully at the behavior of the LLE for the first-order phase transition 
by focusing on values of the parameter in the vicinity of $\Delta=0.5$. 
Results are shown in Fig.~\ref{fig:ec-1order-dynamic}, and are to be compared with those in Fig.~\ref{fig:ec-1order-thermo}
for the analytic estimates.
The results are summarized in Table~\ref{tab:ec-1order}, and show that
the analytical predictions do not agree with the simulations results in these parameter interval.
On the other hand, $\xi$ does increase with $\Delta$, as expected, as shown in Fig.~\ref{fig:beta-vs-delta}. 

\begin{table}[H]
\centering
\begin{tabular}{ccccc}
\hline \hline
$\Delta$ & Theory & $N = 10^4$ & $N = 10^5$ & $N = 10^6$ \\ \hline
0.49     & 0.039  & 0.051  & 0.061  & 0.056 \\
0.50     & 0.040  & 0.064  & 0.058  & 0.069 \\
0.51     & 0.040  & 0.083  & 0.072  & 0.073 \\
0.52     & 0.048  & 0.074  & 0.073  & 0.096 \\ \hline \hline
\end{tabular}
\caption{Critical exponents obtained theoretically and by TM method in the first-order transitions region.}
\label{tab:ec-1order}
\end{table}

\section{Concluding remarks}
\label{S:5}

We showed that the geometric method in 
Refs.~\cite{casetti1995gaussian,casetti2000geometric,casetti1996riemannian,caiani1997geometry,pettinibook}
can be applied to a more general model than the single cosine HMF model considered by Firpo~\cite{firpo1998analytic}.
This required the use of a semi-analytic approach to estimate microcanonical averages of the fluctuations of the curvature along
trajectories (geodesics) in configuration space, which in principle can be extended to other models. 
We also investigated the power law behavior of the LLE close to the different phase transitions of the GHMF model. 
Although the exact value of the LLE obtained from the analytic approach differs from the numerical estimate
from the tangent map method, which also occurs for the cosHMF model~\cite{miranda2018lyapunov}, 
the estimates for the power law exponent for the LLE are in reasonable agreement for second-order phase transitions, 
with the same value $\xi=1/6$ predicted and observed for the HMF model. This is an indication that this may be a universal
exponent, but requires much more investigation. The possibility of using a renormalization group approach is to be considered
in that direction.

For the first-order transition, the numerical estimates differ from the predicted value, 
but  tend to concur as $\Delta$ grows toward the value where the transition becomes second-order.
Simulations with a higher number of particles closer to the critical energy 
might confirm whether predictions are inaccurate in this case.

As a perspective, the present work can be extended to other one-dimensional and higher dimensional models 
in order to verify whether the critical exponent $\xi=1/6$ indeed qualifies 
as a universal critical exponent for second-order phase-transitions in long-range interacting systems.

\section{Acknowledgments}

MFPSJ was financed by CNPq (Brazil). TMRF was partially financed by CNPq (Brazil) under grant no.\ 305842/2017-0.
YE enjoyed the hospitality and support from CIFMC/UnB while starting this work.



\end{document}